%% file: arx.tex
\documentclass[11pt,twoside]{article}
\usepackage{amsmath}
\usepackage{epsf}
\usepackage{epsfig}

\usepackage{graphicx}
\usepackage{color}
\usepackage{latexsym}
\usepackage{amssymb}
\usepackage{a4wide}
\usepackage{xspace}
\usepackage{pslatex}

\newtheorem{lemma}{Lemma}
\newtheorem{cor}[lemma]{Corollary}
\newtheorem{theorem}[lemma]{Theorem}

\begin{document}

\newcommand{\NP}{\ensuremath{\mathcal{N}\mathcal{P}}\xspace}
\newcommand{\proof}{\noindent\textbf{Proof.}\enspace}
\newcommand{\proofsketch}{\noindent\textbf{Proofsketch.}\enspace}
\newcommand{\qed}{\hfill$\Box$}
\newcommand{\bi}{\begin{itemize}}
\newcommand{\ei}{\end  {itemize}}
\newcommand{\bt}{\begin{tabbing}}
\newcommand{\et}{\end  {tabbing}}
\newcommand{\be}{\begin{enumerate}}
\newcommand{\ee}{\end  {enumerate}}
\newcommand{\bl}{\hspace*{2mm}}
\newcommand{\Bl}{\hspace*{5mm}}
\newcommand{\BL}{\hspace*{10mm}}
\newcommand{\sq}{\sqrt{2}}

\title{The One-Round Voronoi Game Replayed}
\author{
  S\'andor P.\ Fekete
   \thanks{Abteilung f\"ur Mathematische Optimierung,
    TU Braunschweig,
    D-38106 Braunschweig, Germany,
    \texttt{s.fekete@tu-bs.de}.}
  \and
  Henk Meijer  \thanks{
    Department of Computer Science
    Queen's University
    Kingston, Ont K7L 3N6
    Canada,
    \texttt{henk@cs.queensu.ca}.
    Partially supported by NSERC.
    This work was done while visiting TU Braunschweig.}
}

\date{}

\maketitle

\markboth{}{}

%%%%%%%%%%%%%%%%%%%%%%%%%%%%%%%%%%%%%%%%%%%%%%%%%%%%%%%%%%%%%%%%%%%%%%%

\begin{abstract}
We consider the one-round Voronoi game, where the first player (``White'',
called ``Wilma'') places a
set of $n$ points in a rectangular area of aspect ratio $\rho \leq 1$,
followed by the second player (``Black'', called ``Barney''), 
who places the same number of points.
Each player wins the fraction of $A$ closest to one of his points,
and the goal is to win more than half of the total area.
This problem has been studied by Cheong et al.\,
who showed that for large enough $n$ and $\rho = 1$, Barney
has a strategy that guarantees a fraction of $1/2+\alpha$, for some
small fixed $\alpha$. 

We resolve a number of open problems raised by that paper.
In particular, we give a precise characterization of the outcome of the game
for optimal play:
We show that Barney has a winning strategy
for $n\geq 3$ and $\rho >\sqrt{2}/n$, and for $n=2$ and $\rho>\sqrt{3}/2$.
Wilma wins in all remaining cases, i.e., for
$n\geq 3$ and $\rho \leq \sqrt{2}/n$, for $n=2$ and $\rho \leq\sqrt{3}/2$,
and for $n=1$. We also discuss complexity aspects of the game on more general
boards, by proving that for a polygon with holes,
it is NP-hard to maximize the area Barney
can win against a given set of points by Wilma.
\end{abstract}

{\bf ACM-Classification}: F.2.2: Nonumerical Algorithms and Problems: Geometrical Problems and Computations

\bigskip
{\bf MSC-Classification}: 90B85, 91A05, 68Q25

\bigskip
{\bf Keywords:}
Computational geometry, Voronoi diagram, Voronoi game, 
competitive facility location, 2-person games, NP-hardness.

\newpage
\section{Introduction}
When determining success or failure of an enterprise,
{\em location} is one of the most important issues.
Probably the most natural way to determine the value
of a possible position for a facility
is the distance to potential customer sites. 
Various geometric scenarios have been considered;
see the extensive list of references in the paper by Fekete,
Mitchell, and Weinbrecht~\cite{FMW00}
for an overview.

One particularly important issue in location theory is the
study of strategies for competing players. See
the surveys by Tobin, Friesz, and Miller~\cite{TFM89},
by Eiselt and Laporte~\cite{EL89}, and by Eiselt, Laporte,
and Thisse~\cite{ELT93}. 

A simple geometric model for the value of a position
is used in the {\em Voronoi game}, which was proposed
by Ahn et al.~\cite{ACCGO.jour} (calling the two-dimensional
scenario the most natural one), and solved for the one-dimensional
scenario. Cheong et al.~\cite{CHLM.jour} provided results
for the two- and higher-dimensional case. In this game, a site $s$ ``owns''
the part of the playing arena that is closer to $s$ than to
any other site. Both considered a two-player
version with a finite arena $Q$. The players, White (``Wilma'')
and Black (``Barney''), place points in $Q$; Wilma plays first.
No point that has been occupied can be changed or reused 
by either player. Let $W$ be the set of points that were
played by the end of the game by Wilma, 
while $B$ is the set of points played by Barney.
At the end of the game, a Voronoi diagram of $W\cup B$ is constructed;
each player wins the total area of all cells belonging to points
in his or her set.  The player with the larger total area wins.

Ahn et al.~\cite{ACCGO.jour} showed that for a one-dimensional
arena, i.e., a 
line segment $[0,2n]$, Barney can win the {\em $n$-round game},
in which each player places a single point in each turn;
however, Wilma can keep Barney's winning margin arbitrarily small.
This differs from  the {\em one-round game}, in which both players
get a single turn with $n$ points each:
Here, Wilma can force a win by playing the odd integer points
$\{1,3,\ldots,2n-1\}$; again, the losing player can make the
margin as small as he wishes. The used strategy focuses on ``key points''.
The question raised in the end of that paper is whether a similar
notion can be extended to the two-dimensional scenario.
We will see in Section~\ref{sec:grid} that in a certain sense,
this is indeed the case.

Cheong et al.~\cite{CHLM.jour} showed that the 
two- or higher-dimensional scenario differs significantly:
For sufficiently large $n\geq n_0$ and a square playing surface $Q$, 
the second player has a winning
strategy that guarantees at least a fixed fraction of
$1/2+\alpha$ of the total area. Their proof uses a clever
combination of probabilistic arguments to show that Barney
will do well by playing a random point. The paper gives
rise to some interesting open questions:

\begin{itemize}
\addtolength{\itemsep}{-2mm}
\item How large does $n_0$ have to be to guarantee a winning
strategy for Barney? 
Wilma wins for $n=1$, but it is not clear
whether there is a single $n_0$ for which the game changes
from Wilma to Barney,
or whether there are multiple changing points.
\item For sufficiently ``fat''
arenas, Barney wins, while Wilma wins for the degenerate case
of a line.
How exactly does the outcome of the game depend on the
aspect ratio of the playing board? 
\item What happens if the number of points played by Wilma
and Barney are not identical?
\item What configurations of white points limit the
possible gain of black points? As candidates, square or
hexagonal grids were named.
\item What happens for the multiple-round version of the game?
\item 
What happens for asymmetric playing boards?
\end{itemize}

For rectangular boards and arbitrary values of $n$, 
we will  give a precise characterization of when Barney can win
the game. If the board $Q$ has aspect ratio $\rho$ with $\rho \leq 1$,
we prove the following:

\begin{itemize}
\item Barney has a winning strategy
for $n\geq 3$ and $\rho >\sqrt{2}/n$, and for $n=2$ and $\rho >\sqrt{3}/2$.
Wilma wins in all remaining cases, i.e., for
$n\geq 3$ and $\rho \leq \sqrt{2}/n$, for $n=2$ and $\rho \leq \sqrt{3}/2$,
and for $n=1$.
\item If Wilma does not play her points on an orthogonal grid,
then Barney wins the game.
\end{itemize}

In addition, we hint at the difficulties of more complex
playing boards by showing the following:

\begin{itemize}
\item If $Q$ is a polygon with holes, and Wilma has made her move,
it is NP-hard to find a position of black points that maximizes
the area that Barney wins.
\end{itemize}

This result is also related to recent work by Dehne, Klein, and Seidel
\cite{DKS02} of a different type: They studied the problem of 
placing a single black point within the convex hull
of a set of white points, such that the resulting
black Voronoi cell in the unbounded Euclidean plane is maximized.
They showed that there is a unique local maximum.
For the problem of finding a location for one additional point among 
$n$ given points on a torus that maximizes the resulting 
largest Voronoi cell, see the more recent paper by Cheong, Efrat, and 
Har-Peled \cite{CEH04}, who give a near-linear polynomial-time approximation
scheme.

The rest of this paper is organized as follows.
After some technical preliminaries in Section~~\ref{sec:prelim},
Section~\ref{sec:grid} shows that Barney always wins if
Wilma does not place her points on a regular orthogonal grid.
This is used in Section~\ref{sec:aspect} to establish
our results on the critical aspect ratios.
Section~\ref{sec:complexity} presents some results on
the computational complexity of playing optimally in a more
complex board. Some concluding thoughts are presented in Section~\ref{sec:conc}.

\section{Preliminaries}
\label{sec:prelim}
In the following, $Q$ is the playing board. 
$Q$ is a rectangle of {\em aspect ratio} $\rho$, which
is the ratio of the length of the smaller
side divided by the length of the longer side.
Unless noted otherwise (in some parts of Section~\ref{sec:complexity}),
both players play $n$ points; $W$ denotes
the $n$ points played by Wilma, while $B$ is the set of $n$ points played
by Barney. All distances are measured according to the Euclidean norm.
For a set of points $P$, we denote by $V(P)$ the (Euclidean)
Voronoi diagram of $P$.
We call a Voronoi diagram $V(P)$ a regular grid if
\bi
\item 
all Voronoi cells are rectangular, congruent and have the same orientation;
\item 
each point $p \in P$ lies in the center of its Voronoi cell.
\ei
If $e$ is a Voronoi edge, $C(e)$ denotes a Voronoi
cell adjacent to $e$.
If $p \in P$, then $C(p)$ denotes the  Voronoi
cell of $p$ in $V(P)$. $\partial C(p)$ is the boundary of $C(p)$ and
$|C(p)|$ denotes the area of  $C(p)$.
$|e|$ denotes the length of an edge $e$.
Let $x_p$ and $y_p$ denote the $x$- and $y$-coordinates of a point $p$.

\section{A Reduction to Grids}
\label{sec:grid}
As a first important step, we reduce the possible configurations
that Wilma may play without losing the game.
The following holds for boards of any shape:

\begin{lemma}
\label{le:symmcell}
If $V(W)$ contains a cell that is not point symmetric, then Barney wins.
\end{lemma}

\proof
Let $r(w,\varphi)$ be the distance from a point $w$ in $C(w)$
to the point on
the boundary of $C(w)$ that is stabbed by a ray emanating from
$w$ at angle $\varphi$. 
Let $l(w,\varphi)$ be the line that contains the ray $r(w,\varphi)$.
As $\partial C(w)$ is a convex curve, $r(w,\varphi)$
is a continuous function. Furthermore, we see that 
$|C(p)|=\frac{1}{2}\int_0^{2\pi}r^2(w,\varphi)d\varphi$, and the portion
of $C(p)$ enclosed between angles $\varphi_1$ and $\varphi_2$ is
$\frac{1}{2}\int_{\varphi_1}^{\varphi_2}r^2(w,\varphi)d\varphi$. 
So an infinitesimal rotation of $l(w,\varphi)$ about $w$ changes
the area by $\pm (r^2(w,\varphi) - r^2(w,-\varphi))d\varphi$.

For all points $w$ in a non-symmetric cell,
there is a  $\varphi$ for which $r(w,\varphi) \neq r(w,-\varphi)$.
Let $w$ be the location in a non-symmetric cell of $V(W)$
where Wilma has placed her point.
Let $\varphi$ be such 
that $r(w,\varphi) \neq r(w,-\varphi)$.
So either the line $l(w,\varphi)$ does not bisect the area of $C(w)$
or we can rotate this line around $w$ so that it does not
bisect the area of $C(w)$.
Therefore there is a line through $w$ such that 
we have an area of size $|C(w)|/2 + 2\varepsilon$ on one side
of this line for some small positive value of $\varepsilon$.  
This means that by placing a point
close to $w$, Barney can claim at least $|C(w)|/2 + 2\varepsilon -\varepsilon/n$
of the cell $C(w)$.
In each other cell $C(w)$ of $V(W)$
Barney can place a point 
close enough to $w$ to claim an area of at least $|C(w)|/2  - \varepsilon /n$.
Therefore Barney has gained at least $|Q|/2 + \varepsilon$.
\qed

\begin{cor}
\label{cor:symmcell}
If all cells of $V(W)$ are point symmetric but Wilma has
not placed placed all her points in the centres of each cell,
then Barney wins.
\end{cor}

\proof
Follows from the fact that the argument used in the proof of Lemma \ref{le:symmcell}
applies whenever a point of Wilma is not placed in the centre of its cell.
\qed

The following theorem is based on this observation and will be used
as a key tool for simplifying our discussion in Section~\ref{sec:aspect}.

\begin{theorem}
\label{th:grid}
If the board is a rectangle and if $V(W)$ is not a regular grid, 
then Barney wins.
\end{theorem}

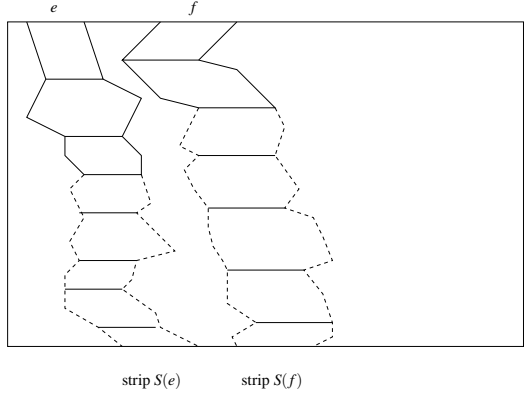
\begin{figure}[htb]
   \begin{center}
   \input{strip.0.pstex_t}
   \caption{Playing board with two strips.}
   \label{fi:strip.0}
   \end{center}
\end{figure}

\proof
We assume that Barney cannot win, and will show that this implies
that $V(W)$ is a regular grid.
By Lemma \ref{le:symmcell},
we may assume that all cells of  $V(W)$ are  point symmetric.
By  Corollary \ref{cor:symmcell} we know that the points in $W$
are the centres of the cells of $V(W)$.
Let $e_0$ be a Voronoi edge of $V(W)$ on the top side 
of the board. 
Consider the Voronoi cell $C_0$ adjacent to $e_0$. 
Because $C_0$ is point symmetric, it contains an edge $e_1$
that is parallel to $e_0$ with  $|e_0| = |e_1|$.
Let $C_1$ be the cell adjacent to and below $e_1$.
It contains an edge  $e_2$ with  $|e_2| = |e_1|$.
Similarly define the cells $C_2,C_3,\ldots$.
So cell $C_i$ lies below $C_{i-1}$. 
Therefore there is a cell
$C_{k-1}$ such that $e_k$ lies on the bottom edge of the board.
We call $ S(e_0) = \{C_0,C_1,C_2,\ldots,C_{k-1}\}$ the strip of $e_0$.
Because $C_i$ is  convex, any horizontal line that intersects 
the board
has an intersection with $S(e_0)$ of length $\geq |e_0|$.
Consider two different Voronoi edges, $e$ and $f$, on the top side
of the board, with their respective strips $S(e)$ and $S(f)$.
Because Voronoi cells
are convex and do not have corners with angles of size $\pi$, these strips
cannot intersect, i.e. do not have a cell in common. 
For an illustration, see Figure \ref{fi:strip.0}.
Let $S$ be the collection of strips of $e$ for all Voronoi edges $e$
of $V(W)$ on the top side of the board.  
The intersection of a line with $S$ has a length which is at least as large 
as the sum of the lengths of the edges along the top side of the board.
Because strips do not intersect, this intersection is exactly as
long as the top side of the board. 
This implies that $S$ covers the whole board and that  any horizontal 
line that intersects the board
has an intersection with $S(e_0)$ of length exactly equal to $|e_0|$.

Let $e$ be the left most edge on the top side of the board.
The left hand side of $S(e)$ is the left hand side of the board.
This implies that each cell in  $S(e)$ 
is a rectangle. By the same argument, each cell in $V(W)$ is a rectangle. 
Let $v$ and $w$ be two points in $W$ such that $C(v)$ and $C(w)$ in $V(W)$
have a horizontal edge $e$ in common. The distance between $v$ and $e$ is the
same as the distance between $w$ and $e$. Because both  $C(v)$ and $C(w)$
are point symmetric and rectangular, and $v$ and $w$ are the centres
of $C(v)$ and $C(w)$ respectively, 
it follows that $C(v)$ and $C(w)$ have the same vertical width.
Similarly, if two cells $C(v)$ and $C(w)$
share a vertical edge, these cells have the same horizontal width.
Therefore $V(W)$ is a regular grid.  \qed

\section{Critical Aspect Ratios}
\label{sec:aspect}
In this section we prove the main result of this paper: 
if $n\geq 3$ and $\rho > \sqrt{2}/n$, or 
$n=2$ and $\rho >\sqrt{3}/2$, then Barney wins.
In all other cases, Wilma wins.
The proof proceeds by a series of lemmas.
We start by noting the following easy observation. 

\begin{lemma}
\label{le:bigger}
Barney wins, if and only if he can place a point $p$ that steals an
area strictly larger than $|Q|/2n$ from $W$.
\end{lemma}

\proof
Necessity is obvious. To see sufficiency, note that 
Wilma is forced to play her points in a regular grid.
Barney places his first point $p$ such that it gains an area of more 
than  $|Q|/2n$.
Let $w$ be a point in $W$. If Barney places a point on the line through $w$ and $p$,
sufficiently close to $w$ but on the opposite side of $p$, he can claim
almost half of the Voronoi cell of $w$.
By placing his remaining $n-1$ points
in this fashion, he can claim an area larger than  $|Q|/2n$.  \qed

Next we take care of the case $n=2$; this lemma will also be useful
for larger $n$, as it allows further reduction of the possible
arrangements Wilma can choose without losing.

\begin{lemma}
\label{le:twostones}
If $n=2$ and
$\rho >\sqrt 3/2$,
then Barney wins. If the aspect ratio is smaller, Barney loses.
\end{lemma}

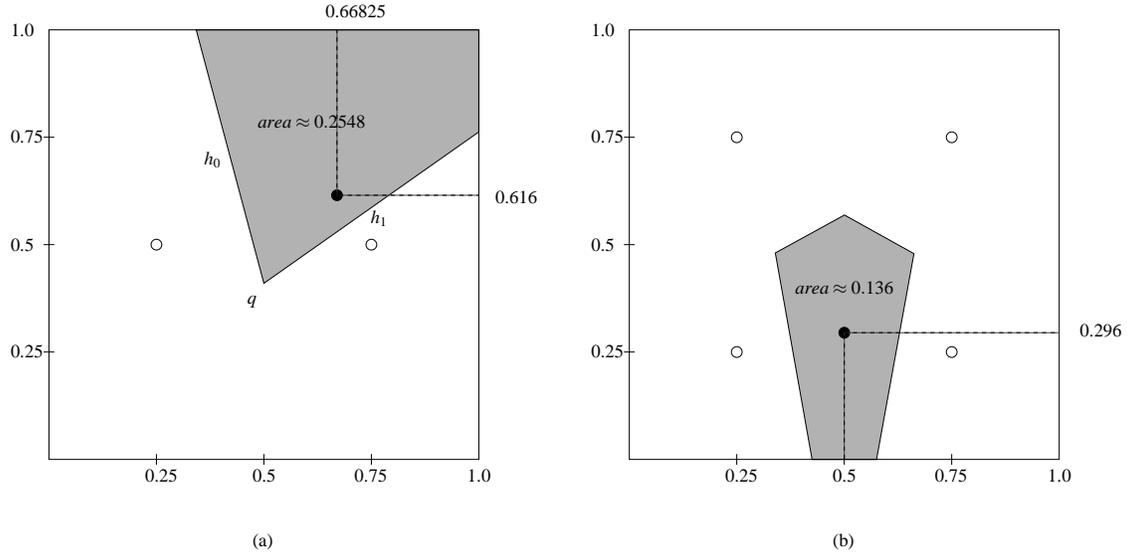
\begin{figure*}[htb]
   \begin{center}
   \input{twostones.0.pstex_t}
   \caption{Barney has gained more than a quarter (a) more than an eighth (b) of the playing surface.}
   \label{fi:twostones.0}
   \end{center}
\end{figure*}

\proof
Assume without loss of generality that 
the board has size $\rho$ by 1.
Suppose that the left bottom corner of $Q$ lies on the origin.
By Theorem~\ref{th:grid} we know that Wilma has to place her
points at $(0.5,\rho/4)$ and $(0.5,3\rho/4)$ or at
$(0.25,\rho/2)$ and $(0.75,\rho/2)$.
If Wilma places her points at $(0.5,\rho/4)$ and $(0.5,3\rho/4)$,
then it is not hard to show that she will lose.
So assume that Wilma places her points at $(0.25,\rho/2)$ and $(0.75,\rho/2)$.
For Barney to win, he will have to gain more than $\rho/4 $
with his first point. Suppose Barney places his point at location $p$.
Without loss of generality, assume that $x_p \geq 0.5$
and $y_p \geq \rho/2$. If  $y_p = \rho/2$
then Barney gains at most  $\rho/4 $, so we may assume that  $y_p > \rho/2$.
Placing a point $p$ with $x_p > 0.75$
is not optimal for Barney: moving $p$ in the direction of
$(0.5,\rho/2)$ will
increase the area gained. It is not hard to show that for 
$x_p=0.75$, Barney cannot gain an area of size $\rho/4$.
So we may assume that $0.5 \leq x_p < 0.75$.
Let $b_0$ be the bisector of $p$ and $(0.25,\rho/2)$. Let $b_1$
be the bisector of $p$ and $(0.75,\rho/2)$. Let $q$ be the intersection of $b_0$
and $b_1$. The point $q$ lies on the vertical line through $x = 0.5$.
If $q$ lies outside the board $Q$, then $|C(p)| < \rho/4$,
so assume that $q$ lies in $Q$.
Let $h_0$ be the length of the line segment
on $b_0$, between $q$ and the top or left side of the board.
Let $h_1$ be the length of the line segment
on $b_1$, between $q$ and the top or right side of the board.
Consider the circle $C$
center at $q$ which passes through $p$, $(0.25,\rho/2)$ and
$(0.75,\rho/2)$. 

If $b_0$ does not intersect the top of the board then neither does
$b_1$.
In this case we can increase $|C(p)|$ by moving $p$
to the left on $C$ and we can
use this to show that $|C(p)| < \rho/4$.
If both $b_0$ and $b_1$ intersect the top of the board we
have $h_0 \leq h_1$. We can increase $h_1$
and decrease $h_0$ by moving $P$ to the right on $C$.
So $|C(p)|$
can be increased until $b_1$ intersects the top right corner
of the board.
If $b_0$ intersect the top of the board
and $b_1$ intersect the right top corner we have
$h_0 \leq h_1$. If we move $p$ to the right on $C$, both $h_0$
and $h_1$ will decrease. The area $|C(p)|$ will increase
as long as $h_0 < h_1$ and reaches its maximum value when $h_0 = h_1$.
Therefore the maximum exists when at the moment that 
$p$ approaches $(0.75,\rho/2)$, 
we have $h_0 > h_1$.
When  $p = (0.75,\rho/2)$, we have 
$h_0 = \rho - y_q$ and $h_1 = \sqrt{(1/4 + (\rho - 2y_q)^2)}$.
From  $h_0 > h_1$ we can derive that $\rho > \sqrt 3 / 2$.
With his second point Barney can gain an area of size $0.25 - \epsilon$ for an arbitrary
small positive value of $\epsilon$ by placing the point
close to $(0.25,\rho/2)$. So Barney can gain more than half the board.

If the aspect ratio is $\leq \sqrt 3/2$, Barney can gain at most $\rho/4$
with his first move by placing his point at $(x,\rho/2)$ with $0.25 < x < 0.75$. 
It can be shown that with his second point
he can gain almost, but not exactly a quarter.  \qed

The gain for Barney is small if $\rho$ is close to $\sqrt 3/2$.
We have performed computer experiments to compute the 
gain for Barney for values of $\rho > \sqrt 3/2$. 
Not surprisingly, the largest gain was for $\rho = 1$.
If the board has size $1\times 1$, Barney can gain an area of
approximately 0.2548 with his first point, by placing it at (0.66825,0.616)
as illustrated in Figure \ref{fi:twostones.0}(a).

\begin{lemma}
\label{le:fourstones}
Suppose that the board is rectangular and that $n=4$.
If Wilma places her points on a regular $2\times 2$ grid,
Barney can gain $50.78 \%$ of the board.
\end{lemma}

\proof
Assume that the board has size $\rho \times 1$.
By Lemma \ref{th:grid} we know that Wilma has to place her
points on the horizontal line at height $\rho/2$,
on the vertical line at $x= 0.5$ or at the points
$(0.25,\rho/4)$, $(0.25,3\rho/4)$, $(0.75,\rho/4)$ and, $(0.75,3\rho/4)$.
If Wilma does not place her points on a line, it can be computed that Barney wins at 
least $\rho(1/8 + 1/128)$ by
placing a point at $(0.5,\rho/4)$. 
In addition Barney can gain a little more than $3\rho/8 -\epsilon$ 
by placing his remaining three points at 
$(0.25-4\epsilon/3,\rho/4)$, $(0.25-4\epsilon/3,3\rho/4)$, and  $(0.75+4\epsilon/3,3\rho/4)$.
So Barney will gain a total area of size 
$ \rho(1/2 + 1/128) - \epsilon$. Because of $1/2 + 1/128 = 50.78125$, the result 
follows.  \qed

The value in the above lemma is not tight.
For example, if Wilma places her point in a $2\times 2$ grid on a square board, we can
compute the area that Barney can gain with his first point.
If Barney places it at (0.5,0.296), he gains
approximately 0.136.
For an illustration, see Figure \ref{fi:twostones.0}(b).
By placing his
remaining three points at 
$(0.25-4\epsilon/3,0.25)$, $(0.25-4\epsilon/3,0.75)$, and  $(0.75+4\epsilon/3,0.75)$
Barney can gain a total area 
of size  
of around $0.511 - \epsilon$ for arbitrary small positive $\epsilon$.
For non-square boards, we have found larger
wins for Barney.
This suggests that Barney can always gain more than $51\%$ of the board 
if Wilma places
her four points in a $2\times 2$ grid.

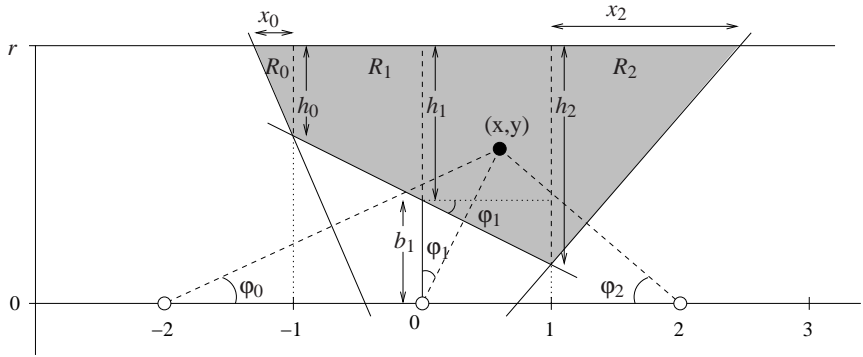
\begin{figure*}[htb]
   \begin{center}
   \input{threestones.pstex_t}
   \caption{Wilma has placed at least three points on a line.}
   \label{fi:threestones}
   \end{center}
\end{figure*}

The above discussion has an important implication:

\begin{cor}
\label{cor:single}
If $n\geq 3$, then Wilma can only win by placing her points
in a $1\times n$ grid.
\end{cor}

This sets the stage for the final lemma:

\begin{lemma}
\label{le:ratio}
Let $n\geq 3$.
Barney can win if $\rho > \sqrt{2}/n$; otherwise, he loses. 
\end{lemma}

\proof
It follows from Corollary~\ref{cor:single}
that Wilma should place her points in a $1\times n$ grid.
Assume that $Q$ has size $2r \times 2n$ and that the left bottom point
of $Q$ lies at $(-3,-r)$ and the top right point at $(2n-3,r)$.
Wilma must place her points at
$(-2,0),(0,0),(2,0),\ldots,(2n-4,0)$.
{From} Lemma \ref{le:bigger} we know that 
in order to win, Barney has to find a location $p=(x,y)$
with $|V(p)|>2r$.

If $r > \sqrt 3$, we know from Lemma \ref{le:twostones} that Barney can take
more than a quarter from two neighboring cells of Wilma, i.e. Barney takes
more than $8r/4 = 2r$ with his first point. Therefore assume that $r \leq \sqrt 3$.
We start by describing the size and area of a potential Voronoi cell
for Barney's first point. 
Without loss of generality, we assume that 
$p=(x,y)$ with $y,x\geq 0$ is placed in the cell of Wilma's
point $(0,0)$, so $x\leq 1$, $y\leq r$.

If $y>0$ and if  Barney gains parts of  three cells of $V(W)$ with his first
point, we have a situation as shown in Figure~\ref{fi:threestones}.
It is not hard to see that he can steal from at most
three cells: $p$ has distance more than 
2 from all cells not neighboring on Wilma's cells $V(-2,0)$ and $V(2,0)$,
which is more than the radius of 
$\sqrt{r^2+1}\leq 2$ of those cells with respect to their center points.
We see that
\begin{eqnarray}
b_1&=&\frac{y}{2}+\frac{x^2}{2y}\label{ym},\\
\tan\varphi_1&=&\frac{x}{y}\label{tanphi},\\
\tan\varphi_2&=&\frac{y}{2-x}\label{tanb+}.
\end{eqnarray}

As shown in Figure~\ref{fi:threestones}, the Voronoi 
cell of $p$ consists of three pieces:
the quadrangle $R_1$ (stolen from $V(0,0)$),
the triangle $R_0$ (stolen from $V(-2,0)$),
and the triangle $R_2$ (stolen from $V(2,0)$).
Furthermore, 
\begin{eqnarray}
|R_1|&=&2 h_1 = 2(r-b_1)=2r-y-\frac{x^2}{y}\label{R1},\\
|R_2|&=&\frac{x_2h_2}{2}\label{R2},  \\
h_2&=&r-b_1+\tan\varphi_1\label{h2}\\
&=&r-\frac{y}{2}-\frac{x^2}{2y}+\frac{x}{y}\label{y+},\\
x_2&=&h_2 \tan\varphi_2=\frac{\left(r-\frac{y}{2}-\frac{x^2}{2y}+\frac{x}{y}\right)y}{2-x}\label{x+}, 
\end{eqnarray}
so
\begin{eqnarray}
|R_2|%&=&\frac{\left(r-\frac{y}{2}-\frac{x^2}{2y}+\frac{x}{y}\right)^2y}{2(2-x)\nonumber}\\
     &=&\frac{\left(r y-\frac{y^2}{2}-\frac{x^2}{2}+x\right)^2}{2y(2-x)}\label{R21}.
\end{eqnarray}
Analogously,
\begin{eqnarray}
|R_0|&=&\frac{\left(r y-\frac{y^2}{2}-\frac{x^2}{2}-x\right)^2}{2y(2+x)}\label{R0}.
\end{eqnarray}
We first consider $r \leq \sqrt 2$. Assume that Barney can win, i.e. can gain
an area larger than $2r$ with his first point.
If $y=0$, then
$|V(p)|=2r$, so we may assume that  $y>0$.
{From} Lemma~\ref{le:twostones}, we
know that Barney will not win if he only steals from two of Wilma's cells, so we may assume
that  Barney steals from three cells. Therefore we can use results from equations (\ref{R1}),
(\ref{R21}) and (\ref{R0}).
From $|R_0| + |R_1|+|R_2|>2r$ we derive
\begin{eqnarray}
&&\left(r y-\frac{y^2}{2}-\frac{x^2}{2}-x\right)^2(2-x)\nonumber\\
&+& \left(r y-\frac{y^2}{2}-\frac{x^2}{2}+x\right)^2(2+x)\nonumber\\
&>& 2(y^2+x^2)(4-x^2) \label{one}  
\end{eqnarray}
As the left hand side is maximized for $r=\sqrt{2}$, we conclude
\begin{eqnarray}
&& \left(\sqrt{2}y-\frac{y^2}{2}-\frac{x^2}{2}-x\right)^2(2-x)\nonumber\\
&+&\left(\sqrt{2}y-\frac{y^2}{2}-\frac{x^2}{2}+x\right)^2(2+x)\nonumber\\
&>& 2(y^2+x^2)(4-x^2), \label{two} 
\end{eqnarray}
so 
\begin{eqnarray}
&&4\left(\sq y-\frac{y^2}{2}-\frac{x^2}{2}\right)^2\nonumber\\
&+&4x^2\left(\sq y-\frac{y^2}{2}-\frac{x^2}{2}\right)+4x^2\nonumber\\
&>& 8y^2+8x^2-2x^2y^2-2x^4, \label{three}
\end{eqnarray}
implying
 \begin{eqnarray}
&&4\left(\left(\sq y-\frac{y^2}{2}-\frac{x^2}{2}\right)+\frac{x^2}{2}\right)^2\nonumber\\
&&-x^4+4x^2\nonumber \\
&>& 8y^2+8x^2-2x^2y^2-2x^4, \label{four} 
\end{eqnarray}
therefore
\begin{eqnarray}
&& 2\left(\sq y-\frac{y^2}{2}\right)^2\nonumber\\
&>& 4y^2+2x^2-\frac{x^4}{2}-x^2y^2 \label{five} 
\end{eqnarray}
and thus
\begin{eqnarray}
&& 4y^2-2\sq y^3+\frac{y^4}{2}\nonumber\\
&>& 4y^2+x^2\left(2-\frac{x^2}{2}-y^2\right) \label{six} 
\end{eqnarray}
or
\begin{eqnarray}
&& y^3(\frac{y}{2}-2\sq)\nonumber\\
&>& x^2\left(2-\frac{x^2}{2}-y^2\right).\label{seven}
\end{eqnarray}
As the left hand side is negative for $0 < y \leq \sqrt 2$, we conclude that
the right hand side must also be negative; clearly, it is minimized
for $x=1$, so we get
\begin{eqnarray}
y^3(\frac{y}{2}-2\sq)
&>&\left(2-\frac{1}{2}-y^2\right),\label{eight}
\end{eqnarray}
and conclude that $\sqrt 2 \geq y \geq \sqrt{3/2}$,
yielding the contradiction
\begin{eqnarray}
4~\geq~ \frac{y^4}{2}+y^2
&>&\frac{3}{2}+2\sq y^3~>~4.\label{nine}
\end{eqnarray}

So the best Barney can do is gain an area of size  $2r$
with all his points and tie the game.
However, notice that the contradiction in equation (\ref{nine}) also holds if
$|R_0| + |R_1|+|R_2|=2r$. So Barney cannot gain an area of size $2r$
if he  places his point at $(x,y)$ with $y>0$
and steals from three cells of $V(W)$.
In Lemma \ref{le:twostones} it was shown that 
Barney will gain less than $2r$ if he places his point at $(x,y)$ with $y>0$
and  steals from two cells of $V(W)$.
Therefore Barney must place his points at $(x,y)$ with $y=0$.
This reduces the problem to a one-dimensional one, and we
know from  \cite{ACCGO.jour} that in that case Barney will lose. 

Secondly we consider $\sqrt 2 < r \leq \sqrt 3$. 
Suppose Barney places his first point at $(0,y)$ with $y>0$. Clearly he will steal
from three cells of $V(W)$.
{From} equations  (\ref{R1}), (\ref{R21}) and (\ref{R0}) we derive that
\begin{eqnarray}
&&|R_0|  + |R_1|+|R_2|\nonumber\\
 &=& \frac{r^2 y}{2} - \frac{r y^2}{2} +  \frac{y^3}{8} + 2r - y.  \label{ten}
\end{eqnarray}
Because of $y > 0$ we have
\begin{eqnarray}
&& |R_0|  + |R_1|+|R_2| ~>~  2r \\
&\Leftrightarrow& \frac{r^2 y}{2} - \frac{r y^2}{2} +  \frac{y^3}{8} - y ~>~ 0   \\
&\Leftrightarrow& y^2 - 4 r y + 4 r^2 - 8 ~>~ 0   \\
&\Leftrightarrow& 0 < y < 2 (r - \sqrt 2).
\end{eqnarray}
So Barney wins if he places
a point at $(0,y)$ with $0 < y < 2(r - \sqrt 2)$.  \qed

\bigskip
The value in equation (\ref{ten}), i.e., the total 
area, is maximal for $y^* = (4 r - 2\sqrt{r^2+6})/3$.
Computational experiments have confirmed that  
Barney maximizes the area with his first 
point at $(0,y^*)$.

Summarizing, we get:

\begin{theorem}
\label{th:main}
If $n\geq 3$ and $\rho > \sqrt{2}/n$, or 
$n=2$ and $\rho >\sqrt{3}/2$, then Barney wins.
In all other cases, Wilma wins.
\end{theorem}

%\proof
%Follows from Lemmas \ref{le:twostones} and  \ref{le:ratio}.  \qed

\section{A Complexity Result}
\label{sec:complexity}

The previous section resolves most of the questions for the one-round
Voronoi game on a rectangular board. Clearly, there are various
other questions related to more complex boards; this is
one of the questions raised in \cite{CHLM.jour}. Lemma~\ref{le:symmcell}
still applies if Wilma's concern is only to avoid a loss.
Moreover, it is easily seen
that all of Wilma's Voronoi cells must have the
same area, as Barney can steal almost all the area of the largest cell
by placing two points in it, and no point in the smallest cell.
For many boards, both of these conditions may be impossible
to fulfill. It is therefore natural to modify the game by shifting
the critical margin that decides a win or a loss.
We show in the following that it is NP-hard to decide whether
Barney can beat a given margin for a polygon with holes,
and all of Wilma's points have already been placed.
(In a non-convex polygon, possibly with holes, we measure distances 
according to the geodesic Euclidean metric, i.e., along a shortest path
within the polygon.)

\begin{theorem}
\label{th:npc}
For a polygon with holes, it is NP-hard to maximize the area Barney can claim,
even if all of Wilma's points have been placed.
\end{theorem}

\begin{figure}[htb]
 \begin{center}
  \leavevmode
  \epsfig{file=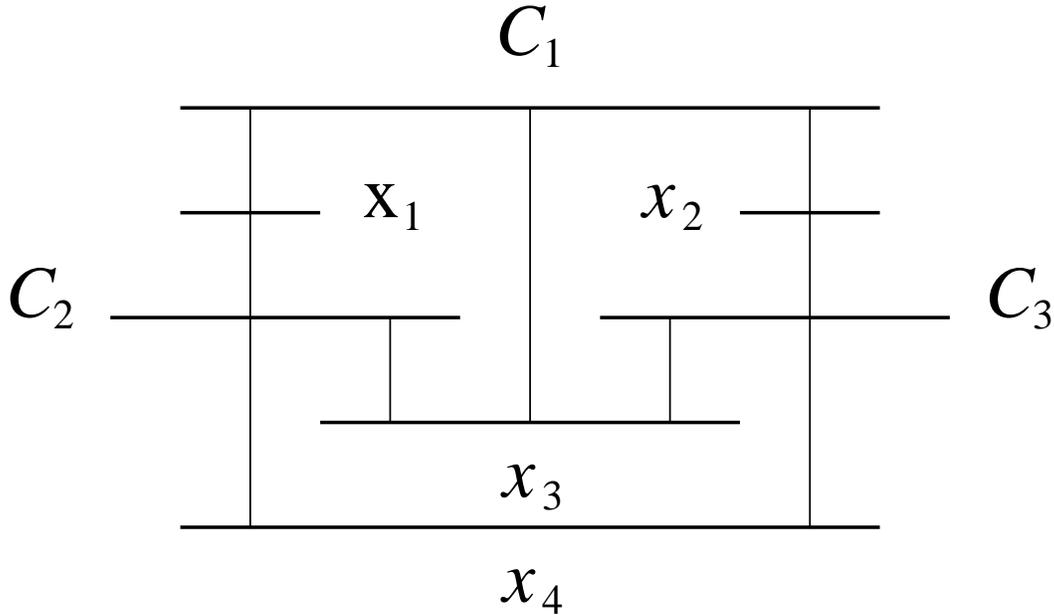, width=0.9\columnwidth}
  \caption{A geometric representation of the graph $G_I$ for the 
    Planar 3SAT instance
  $I=(x_{1} \vee x_{2} \vee x_{3}) 
\wedge (\bar x_{1} \vee \bar x_{3} \vee \bar x_{4}) 
\wedge (\bar x_{2} \vee \bar x_{3} \vee x_{4})$.}
  \label{fig:planar-3-sat}
 \end{center}
\end{figure}

\proof
We give an outline of the proof, based on
a reduction from {\sc Planar 3SAT}, which is known to be 
NP-complete \cite{Lic82}. 
For clearer description, we sketch the proof for the case
where Barney has fewer points to play; in the end,
we hint at what can be done to make both point sets the same size.
(A 3SAT instance $I$ 
is said to be an instance of {\sc Planar 3SAT},
if the following bipartite graph $G_I$ is planar:
every variable $x_i$ and every clause $c_j$ in $I$ is represented
by a vertex in $G_I$; two vertices are connected, if and only
if one of them represents a variable that appears in the
clause that is represented by the other vertex.)
First, the planar graph corresponding to an instance $I$ of {\sc
Planar 3Sat} is represented geometrically as a {\em planar rectilinear
layout}, with each vertex corresponding to a horizontal line segment,
and each edge corresponding to a vertical line segment that
intersects precisely the line segments corresponding to the two
incident vertices. There are well-known algorithms (e.g.,
\cite{rose_tar}) that can achieve such a layout in linear time and
linear space.  See Figure~\ref{fig:planar-3-sat}.

\begin{figure*}[htb]
 \begin{center}
  \leavevmode
  \centerline{\epsfig{file=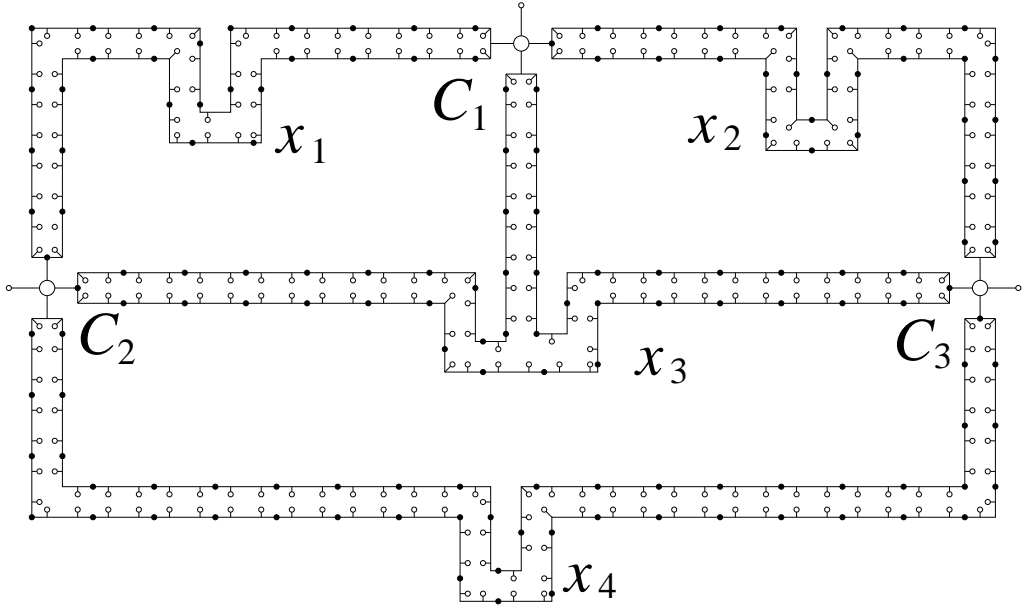, width=.87\textwidth}}
  \caption{A symbolic picture of the overall representation: The location of white points
is indicated by white dots (with area elements on variable loops not drawn
for the sake of clarity). The location of black points (indicated by black dots)
corresponds to the truth assignment $x_1=0$, $x_2=1$, $x_3=0$, $x_4=1$,
which satisfies $I$. See Figure~\ref{fig:gadget} for a closeup of the gadgets.}
  \label{fig:truth}
 \end{center}
\end{figure*}

Next, the layout is modified such that the line segments corresponding
to a vertex representing a literal and all edges incident to it are 
replaced by a loop -- see
Figure~\ref{fig:truth}.  At each vertex
corresponding to a clause, three of these loops (corresponding to the
respective literals) meet.  Each loop gets represented by a very
narrow corridor. 

\begin{figure*}[htb]
   \begin{center}
   \input{gadget.pstex_t}
  \caption{Area gadget (left) and clause gadgets (right)}
  \label{fig:gadget}
   \end{center}
\end{figure*}
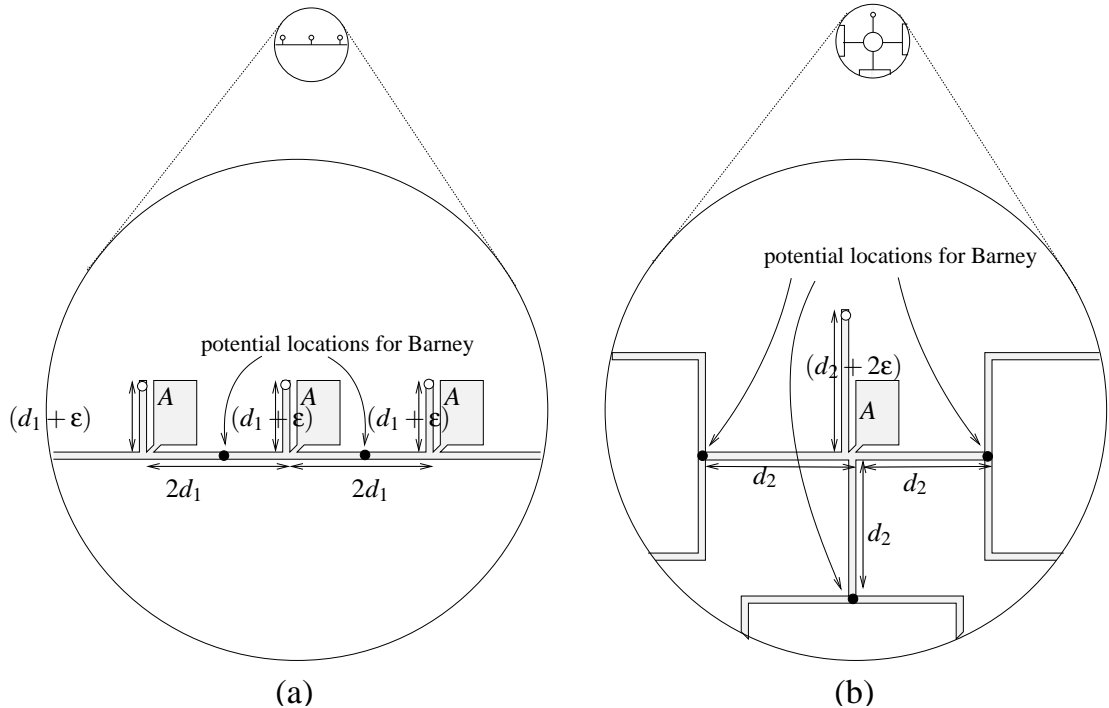

Now we place a sequence of extra
{\em area gadgets} at equal distances $2d_1$ along the variable loop.
Let $n_i$ be the number of area gadgets along the loop for 
variable $x_i$, and let $N=\sum_{i=1}^n n_i$, and $\varepsilon=1/N^3$. 
(By construction, $N$ is polynomially bounded.)
As shown in Figure~\ref{fig:gadget}(a), each 
each such gadget consists of an
area element of size $A=1/N$, ``guarded'' by 
a white point that is
at distance $d_1+\varepsilon$ from it.
Finally, for each clause, we place an extra gadget as shown in 
Figure~\ref{fig:gadget}(b). 
Similar to the area gadgets along the variable loops,
it consists of a white point guarding
an area element of size $A=1/N$ at distance $d_2+\varepsilon$. 
Thus, the overall number of white points is $|W|=N+m$.
By making the corridors sufficiently narrow (say, $1/N^3$ wide),
the overall area for the corridors is small (e.g., $O(1/N^2)$.)
The total area of the resulting polygon is $1+m/N+O(1/N^2)$.
See Figure~\ref{fig:truth} for a symbolic overall picture.

As indicated in Figure~\ref{fig:gadget}, there is
a limited set of positions where a black point can
steal more than one area gadget. Stealing all
area gadgets along a variable loop is possible with $n_i/2$
points, by picking
every other potential location along the loop.
This can be done in two ways, and either such choice represents 
a truth assignment of the corresponding
variable. In a truth assignment in which the variable satisfies
a clause, a black point is placed on the variable loop in such 
a manner that the area element of the clause gadget is stolen.
(See Figure~\ref{fig:truth} for our example.) 
Thus, a satisfying truth assignment
for $I$ yields a position of $N/2$ black points that steals
all the area elements, i.e., claims an area of $1+m/N$.

To see the converse, assume that Barney can claim an area of at least
$1+m/N$, i.e., he can steal all area elements. 
As noted before, no position of a black point can
steal more than two area elements on a variable; stealing two
requires placing it at less than distance $d_1+\varepsilon$ from
both of them.  As the $N/2$ black
points must form a perfect matching of the $N$ area elements, we conclude
that there are only two basic ways to cover all area elements of
a variable $x_i$ by not more than $n_i/2$ black points,
where each location may be subject to variations of size
$O(\varepsilon)$. One of these perfect matchings
corresponds to setting $x_i$ to {\em true}, the other to {\em false}.  
If this truth assignment
can be done in a way that also steals all area elements of 
clause gadgets, we must have a satisfying truth assignment.  
\qed

By adding some extra area elements (say, of size $3A$) right next to
$N/2+m$ of the white points along variable gadgets, and increasing
$|B|$ to $N+m$, we can modify 
the proof to apply to the case in which $|W|=|B|$.
Similarly, it is straighforward to shift the critical threshold
such that Wilma is guaranteed a constant fraction of the board.

\section{Conclusion}
\label{sec:conc}

We have resolved a number of open problems dealing with the
one-round Voronoi game. There are still several issues that remain
open. What can be said about achieving a fixed margin of win
in all of the cases where Barney can win? We believe that our above techniques
can be used to resolve this issue. As we can already quantify this
margin if Wilma plays a grid, what is still needed is a refined
version of Lemma~\ref{le:symmcell} and Theorem~\ref{th:grid}
that guarantees a fixed margin as a function of the amount that
Wilma deviates from a grid. Eventually, the guaranteed margin should
be a function of the aspect ratio. Along similar lines, we believe
that it is possible to resolve the question stated by \cite{CHLM.jour}
on the scenario where the number of points played is not equal.

There are some real-life situations
where explicit zoning laws enforce a minimum distance between points;
obviously, our results still apply for the limiting case. It seems
clear that Barney will be at a serious disadvantage
when this lower bound is raised, but we leave it to future research to have
a close look at these types of questions.

The most tantalizing problems deal with the multiple-round game.
Given that finding an optimal set of points for a single
player is NP-hard, it is natural to conjecture that the two-player,
multiple round game is PSPACE-hard.
Clearly, there is some similarity to the game of Go on
an $n\times n$ board, which is known
to be PSPACE-hard \cite{LS80} and even EXPTIME-complete \cite{Rob83}
for certain rules. 

However, some of this difficulty results from the possibility
of capturing pieces. It is conceivable that at least for relative
simple (i.e., rectangular) boards, there are less involved
winning strategies. Our results from Section~\ref{sec:aspect}
show that for the cases where Wilma has a winning strategy,
Barney cannot prevent this by any probabilistic or greedy approach:
Unless he blocks one of Wilma's key points by placing a point
there himself (which has probability zero
for random strategies, and will not happen for simple greedy strategies), 
she can simply play those points like in the one-round game and claim
a win. Thus, analyzing these key points may indeed be the key to understanding
the game.

\bibliographystyle{plain}
\bibliography{paper}

\end{document}

%% file: strip.0.pstex_t
\begin{picture}(0,0)%
\includegraphics{strip.0.pstex}%
\end{picture}%
\setlength{\unitlength}{1579sp}%
\begingroup\makeatletter\ifx\SetFigFont\undefined%
\gdef\SetFigFont#1#2#3#4#5{%
  \reset@font\fontsize{#1}{#2pt}%
  \fontfamily{#3}\fontseries{#4}\fontshape{#5}%
  \selectfont}%
\fi\endgroup%
\begin{picture}(8124,6090)(1489,-6733)
\put(3301,-6661){\makebox(0,0)[lb]{\smash{\SetFigFont{6}{7.2}{\rmdefault}{\mddefault}{\updefault}{\color[rgb]{0,0,0}strip $S(e)$}%
}}}
\put(2176,-811){\makebox(0,0)[lb]{\smash{\SetFigFont{6}{7.2}{\rmdefault}{\mddefault}{\updefault}{\color[rgb]{0,0,0}$e$}%
}}}
\put(4351,-811){\makebox(0,0)[lb]{\smash{\SetFigFont{6}{7.2}{\rmdefault}{\mddefault}{\updefault}{\color[rgb]{0,0,0}$f$}%
}}}
\put(5176,-6661){\makebox(0,0)[lb]{\smash{\SetFigFont{6}{7.2}{\rmdefault}{\mddefault}{\updefault}{\color[rgb]{0,0,0}strip $S(f)$}%
}}}
\end{picture}

%% file: twostones.0.pstex_t
\begin{picture}(0,0)%
\includegraphics{twostones.0.pstex}%
\end{picture}%
\setlength{\unitlength}{1776sp}%
\begingroup\makeatletter\ifx\SetFigFont\undefined%
\gdef\SetFigFont#1#2#3#4#5{%
  \reset@font\fontsize{#1}{#2pt}%
  \fontfamily{#3}\fontseries{#4}\fontshape{#5}%
  \selectfont}%
\fi\endgroup%
\begin{picture}(14925,7545)(676,-7561)
\put(2551,-6661){\makebox(0,0)[lb]{\smash{\SetFigFont{7}{8.4}{\rmdefault}{\mddefault}{\updefault}{\color[rgb]{0,0,0}0.25}%
}}}
\put(4051,-6661){\makebox(0,0)[lb]{\smash{\SetFigFont{7}{8.4}{\rmdefault}{\mddefault}{\updefault}{\color[rgb]{0,0,0}0.5}%
}}}
\put(5476,-6661){\makebox(0,0)[lb]{\smash{\SetFigFont{7}{8.4}{\rmdefault}{\mddefault}{\updefault}{\color[rgb]{0,0,0}0.75}%
}}}
\put(7051,-6661){\makebox(0,0)[lb]{\smash{\SetFigFont{7}{8.4}{\rmdefault}{\mddefault}{\updefault}{\color[rgb]{0,0,0}1.0}%
}}}
\put(676,-4936){\makebox(0,0)[lb]{\smash{\SetFigFont{7}{8.4}{\rmdefault}{\mddefault}{\updefault}{\color[rgb]{0,0,0}0.25}%
}}}
\put(676,-3436){\makebox(0,0)[lb]{\smash{\SetFigFont{7}{8.4}{\rmdefault}{\mddefault}{\updefault}{\color[rgb]{0,0,0}0.5}%
}}}
\put(676,-1936){\makebox(0,0)[lb]{\smash{\SetFigFont{7}{8.4}{\rmdefault}{\mddefault}{\updefault}{\color[rgb]{0,0,0}0.75}%
}}}
\put(676,-436){\makebox(0,0)[lb]{\smash{\SetFigFont{7}{8.4}{\rmdefault}{\mddefault}{\updefault}{\color[rgb]{0,0,0}1.0}%
}}}
\put(4126,-1726){\makebox(0,0)[lb]{\smash{\SetFigFont{7}{8.4}{\rmdefault}{\mddefault}{\updefault}{\color[rgb]{0,0,0}$area \approx 0.2548$}%
}}}
\put(7441,-2776){\makebox(0,0)[lb]{\smash{\SetFigFont{7}{8.4}{\rmdefault}{\mddefault}{\updefault}{\color[rgb]{0,0,0}0.616}%
}}}
\put(5071,-181){\makebox(0,0)[lb]{\smash{\SetFigFont{7}{8.4}{\rmdefault}{\mddefault}{\updefault}{\color[rgb]{0,0,0}0.66825}%
}}}
\put(10651,-6661){\makebox(0,0)[lb]{\smash{\SetFigFont{7}{8.4}{\rmdefault}{\mddefault}{\updefault}{\color[rgb]{0,0,0}0.25}%
}}}
\put(12151,-6661){\makebox(0,0)[lb]{\smash{\SetFigFont{7}{8.4}{\rmdefault}{\mddefault}{\updefault}{\color[rgb]{0,0,0}0.5}%
}}}
\put(13576,-6661){\makebox(0,0)[lb]{\smash{\SetFigFont{7}{8.4}{\rmdefault}{\mddefault}{\updefault}{\color[rgb]{0,0,0}0.75}%
}}}
\put(15151,-6661){\makebox(0,0)[lb]{\smash{\SetFigFont{7}{8.4}{\rmdefault}{\mddefault}{\updefault}{\color[rgb]{0,0,0}1.0}%
}}}
\put(8776,-4936){\makebox(0,0)[lb]{\smash{\SetFigFont{7}{8.4}{\rmdefault}{\mddefault}{\updefault}{\color[rgb]{0,0,0}0.25}%
}}}
\put(8776,-3436){\makebox(0,0)[lb]{\smash{\SetFigFont{7}{8.4}{\rmdefault}{\mddefault}{\updefault}{\color[rgb]{0,0,0}0.5}%
}}}
\put(8776,-1936){\makebox(0,0)[lb]{\smash{\SetFigFont{7}{8.4}{\rmdefault}{\mddefault}{\updefault}{\color[rgb]{0,0,0}0.75}%
}}}
\put(8776,-436){\makebox(0,0)[lb]{\smash{\SetFigFont{7}{8.4}{\rmdefault}{\mddefault}{\updefault}{\color[rgb]{0,0,0}1.0}%
}}}
\put(15601,-4636){\makebox(0,0)[lb]{\smash{\SetFigFont{7}{8.4}{\rmdefault}{\mddefault}{\updefault}{\color[rgb]{0,0,0}0.296}%
}}}
\put(4051,-7561){\makebox(0,0)[lb]{\smash{\SetFigFont{7}{8.4}{\rmdefault}{\mddefault}{\updefault}{\color[rgb]{0,0,0}(a)}%
}}}
\put(12151,-7561){\makebox(0,0)[lb]{\smash{\SetFigFont{7}{8.4}{\rmdefault}{\mddefault}{\updefault}{\color[rgb]{0,0,0}(b)}%
}}}
\put(3376,-2236){\makebox(0,0)[lb]{\smash{\SetFigFont{7}{8.4}{\rmdefault}{\mddefault}{\updefault}{\color[rgb]{0,0,0}$h_0$}%
}}}
\put(11626,-4036){\makebox(0,0)[lb]{\smash{\SetFigFont{7}{8.4}{\rmdefault}{\mddefault}{\updefault}{\color[rgb]{0,0,0}$area \approx 0.136$}%
}}}
\put(3976,-4186){\makebox(0,0)[lb]{\smash{\SetFigFont{7}{8.4}{\rmdefault}{\mddefault}{\updefault}{\color[rgb]{0,0,0}$q$}%
}}}
\put(5701,-3061){\makebox(0,0)[lb]{\smash{\SetFigFont{7}{8.4}{\rmdefault}{\mddefault}{\updefault}{\color[rgb]{0,0,0}$h_1$}%
}}}
\end{picture}

%% file: threestones.pstex_t
\begin{picture}(0,0)%
\includegraphics{threestones.pstex}%
\end{picture}%
\setlength{\unitlength}{2131sp}%
\begingroup\makeatletter\ifx\SetFigFont\undefined%
\gdef\SetFigFont#1#2#3#4#5{%
  \reset@font\fontsize{#1}{#2pt}%
  \fontfamily{#3}\fontseries{#4}\fontshape{#5}%
  \selectfont}%
\fi\endgroup%
\begin{picture}(9912,4197)(1201,-4873)
\put(1201,-1336){\makebox(0,0)[lb]{\smash{\SetFigFont{9}{10.8}{\rmdefault}{\mddefault}{\updefault}$r$}}}
\put(6751,-2236){\makebox(0,0)[lb]{\smash{\SetFigFont{9}{10.8}{\rmdefault}{\mddefault}{\updefault}(x,y)}}}
\put(4201,-1561){\makebox(0,0)[lb]{\smash{\SetFigFont{9}{10.8}{\rmdefault}{\mddefault}{\updefault}$R_0$}}}
\put(8251,-1561){\makebox(0,0)[lb]{\smash{\SetFigFont{9}{10.8}{\rmdefault}{\mddefault}{\updefault}$R_2$}}}
\put(7576,-2011){\makebox(0,0)[lb]{\smash{\SetFigFont{9}{10.8}{\rmdefault}{\mddefault}{\updefault}$h_2$}}}
\put(6076,-2011){\makebox(0,0)[lb]{\smash{\SetFigFont{9}{10.8}{\rmdefault}{\mddefault}{\updefault}$h_1$}}}
\put(4126,-961){\makebox(0,0)[lb]{\smash{\SetFigFont{9}{10.8}{\rmdefault}{\mddefault}{\updefault}$x_0$}}}
\put(8176,-886){\makebox(0,0)[lb]{\smash{\SetFigFont{9}{10.8}{\rmdefault}{\mddefault}{\updefault}$x_2$}}}
\put(4576,-2011){\makebox(0,0)[lb]{\smash{\SetFigFont{9}{10.8}{\rmdefault}{\mddefault}{\updefault}$h_0$}}}
\put(8101,-4111){\makebox(0,0)[lb]{\smash{\SetFigFont{9}{10.8}{\rmdefault}{\mddefault}{\updefault}$\varphi_2$}}}
\put(6076,-3661){\makebox(0,0)[lb]{\smash{\SetFigFont{9}{10.8}{\rmdefault}{\mddefault}{\updefault}$\varphi_1$}}}
\put(6676,-3286){\makebox(0,0)[lb]{\smash{\SetFigFont{9}{10.8}{\rmdefault}{\mddefault}{\updefault}$\varphi_1$}}}
\put(5401,-1561){\makebox(0,0)[lb]{\smash{\SetFigFont{9}{10.8}{\rmdefault}{\mddefault}{\updefault}$R_1$}}}
\put(5701,-3586){\makebox(0,0)[lb]{\smash{\SetFigFont{9}{10.8}{\rmdefault}{\mddefault}{\updefault}$b_1$}}}
\put(3901,-4111){\makebox(0,0)[lb]{\smash{\SetFigFont{9}{10.8}{\rmdefault}{\mddefault}{\updefault}$\varphi_0$}}}
\end{picture}

%% file: gadget.pstex_t
\begin{picture}(0,0)%
\includegraphics{gadget.pstex}%
\end{picture}%
\setlength{\unitlength}{2368sp}%
\begingroup\makeatletter\ifx\SetFigFont\undefined%
\gdef\SetFigFont#1#2#3#4#5{%
  \reset@font\fontsize{#1}{#2pt}%
  \fontfamily{#3}\fontseries{#4}\fontshape{#5}%
  \selectfont}%
\fi\endgroup%
\begin{picture}(11483,7331)(-224,-6586)
\put(3376,-4403){\makebox(0,0)[lb]{\smash{\SetFigFont{10}{12.0}{\rmdefault}{\mddefault}{\updefault}{\color[rgb]{0,0,0}$2d_1$}%
}}}
\put(1426,-4403){\makebox(0,0)[lb]{\smash{\SetFigFont{10}{12.0}{\rmdefault}{\mddefault}{\updefault}{\color[rgb]{0,0,0}$2d_1$}%
}}}
\put(7576,-4261){\makebox(0,0)[lb]{\smash{\SetFigFont{10}{12.0}{\rmdefault}{\mddefault}{\updefault}{\color[rgb]{0,0,0}$d_2$}%
}}}
\put(8776,-4861){\makebox(0,0)[lb]{\smash{\SetFigFont{10}{12.0}{\rmdefault}{\mddefault}{\updefault}{\color[rgb]{0,0,0}$d_2$}%
}}}
\put(9151,-4336){\makebox(0,0)[lb]{\smash{\SetFigFont{10}{12.0}{\rmdefault}{\mddefault}{\updefault}{\color[rgb]{0,0,0}$d_2$}%
}}}
\put(8701,-3586){\makebox(0,0)[lb]{\smash{\SetFigFont{10}{12.0}{\rmdefault}{\mddefault}{\updefault}{\color[rgb]{0,0,0}$A$}%
}}}
\put(-224,-3661){\makebox(0,0)[lb]{\smash{\SetFigFont{10}{12.0}{\rmdefault}{\mddefault}{\updefault}{\color[rgb]{0,0,0}$(d_1+\varepsilon)$}%
}}}
\put(2101,-3661){\makebox(0,0)[lb]{\smash{\SetFigFont{10}{12.0}{\rmdefault}{\mddefault}{\updefault}{\color[rgb]{0,0,0}$(d_1+\varepsilon)$}%
}}}
\put(3526,-3661){\makebox(0,0)[lb]{\smash{\SetFigFont{10}{12.0}{\rmdefault}{\mddefault}{\updefault}{\color[rgb]{0,0,0}$(d_1+\varepsilon)$}%
}}}
\put(8116,-3112){\makebox(0,0)[lb]{\smash{\SetFigFont{10}{12.0}{\rmdefault}{\mddefault}{\updefault}{\color[rgb]{0,0,0}$(d_2+2\varepsilon)$}%
}}}
\put(1351,-3436){\makebox(0,0)[lb]{\smash{\SetFigFont{10}{12.0}{\rmdefault}{\mddefault}{\updefault}{\color[rgb]{0,0,0}$A$}%
}}}
\put(2851,-3436){\makebox(0,0)[lb]{\smash{\SetFigFont{10}{12.0}{\rmdefault}{\mddefault}{\updefault}{\color[rgb]{0,0,0}$A$}%
}}}
\put(4351,-3436){\makebox(0,0)[lb]{\smash{\SetFigFont{10}{12.0}{\rmdefault}{\mddefault}{\updefault}{\color[rgb]{0,0,0}$A$}%
}}}
\end{picture}